\documentclass{emulateapj-rtx4}
\usepackage{lscape}

\submitted{Accepted for Publication in the Astronomical Journal: December 11,
2014}
\shorttitle{[Na/Fe] in 47 Tucanae AGB Stars}
\shortauthors{Johnson et al.}

\newcommand\iso[2]{$^{\rm #1}$#2}

\begin{document}

\title{AGB SODIUM ABUNDANCES IN THE GLOBULAR CLUSTER 47 TUCANAE (NGC 104)}

\author{
Christian I. Johnson\altaffilmark{1,2},
Iain McDonald\altaffilmark{3},
Catherine A. Pilachowski\altaffilmark{4},
Mario Mateo\altaffilmark{5},
John I. Bailey, III\altaffilmark{5},
Maria J. Cordero\altaffilmark{6},
Albert A. Zijlstra\altaffilmark{3},
Jeffrey D. Crane\altaffilmark{7},
Edward Olszewski\altaffilmark{8}
Stephen A. Shectman\altaffilmark{7}, and
Ian Thompson\altaffilmark{7}
}

\altaffiltext{1}{Harvard--Smithsonian Center for Astrophysics, 60 Garden
Street, MS--15, Cambridge, MA 02138, USA; cjohnson@cfa.harvard.edu}

\altaffiltext{2}{Clay Fellow}

\altaffiltext{3}{Jodrell Bank Centre for Astrophysics, Alan Turing Building, 
Manchester M13 9PL; iain.mcdonald-2@manchester.ac.uk; 
albert.zijlstra@manchester.ac.uk}

\altaffiltext{4}{Astronomy Department, Indiana University Bloomington, Swain
West 319, 727 East 3rd Street, Bloomington, IN 47405--7105, USA;
catyp@astro.indiana.edu}

\altaffiltext{5}{Department of Astronomy, University of Michigan, Ann Arbor, 
MI 48109, USA; mmateo@umich.edu; baileyji@umich.edu}

\altaffiltext{6}{Zentrum f\"{u}r Astronomie der Universit\"{a}t
Heidelberg, Landessternwarte, K\"{o}nigstuhl 12, Heidelberg, Germany;
mjcorde@lsw.uni-heidelberg.de}

\altaffiltext{7}{Observatories of the Carnegie Institution of Washington, 
Pasadena, CA 91101, USA; crane@obs.carnegiescience.edu; 
ian@obs.carnegiescience.edu; shec@obs.carnegiescience.edu}

\altaffiltext{8}{Steward Observatory, The University of Arizona, Tucson, AZ 
85721, USA ; eolszewski@as.arizona.edu}

\begin{abstract}

A recent analysis comparing the [Na/Fe] distributions of red giant branch (RGB)
and asymptotic giant branch (AGB) stars in the Galactic globular cluster
NGC 6752 found that the ratio of Na--poor to Na--rich stars changes from 30:70
on the RGB to 100:0 on the AGB.  The surprising paucity of Na--rich stars on 
the AGB in NGC 6752 warrants additional investigations to determine if the 
failure of a significant fraction of stars to ascend the AGB is an attribute
common to all globular clusters.  Therefore, we present radial velocities, 
[Fe/H], and [Na/Fe] abundances for 35 AGB stars in the Galactic globular 
cluster 47 Tucanae (47 Tuc; NGC 104), and compare the AGB [Na/Fe] distribution 
with a similar RGB sample published previously.  The abundances and velocities 
were derived from high resolution spectra obtained with the Michigan/Magellan 
Fiber System (M2FS) and MSpec spectrograph on the Magellan--Clay 6.5m 
telescope.  We find the average heliocentric radial velocity and [Fe/H] values 
to be $\langle$RV$_{\rm helio.}$$\rangle$=--18.56 km s$^{\rm -1}$ 
($\sigma$=10.21 km s$^{\rm -1}$) and $\langle$[Fe/H]$\rangle$=--0.68 
($\sigma$=0.08), respectively, in agreement with previous literature estimates.
The average [Na/Fe] abundance is 0.12 dex lower in the 47 Tuc AGB sample 
compared to the RGB sample, and the ratio of Na--poor to Na--rich stars is
63:37 on the AGB and 45:55 on the RGB.  However, in contrast to NGC 6752, the 
two 47 Tuc populations have nearly identical [Na/Fe] dispersion and 
interquartile range values.  The data presented here suggest that only a small 
fraction ($\la$20$\%$) of Na--rich stars in 47 Tuc may fail to ascend the AGB,
which is a similar result to that observed in M13.  Regardless of the cause 
for the lower average [Na/Fe] abundance in AGB stars, we find that Na--poor 
stars and at least some Na--rich stars in 47 Tuc evolve through the early AGB 
phase.  The contrasting behavior of Na--rich stars in 47 Tuc and NGC 6752 
suggests that the RGB [Na/Fe] abundance alone is insufficient for predicting 
if a star will ascend the AGB.

\end{abstract}


\section{INTRODUCTION}

Clear evidence indicates that most Galactic globular clusters host multiple, 
distinct stellar populations (e.g., see reviews by Gratton et 
al. 2004; 2012).  For mono--metallic clusters exhibiting negligible spreads in 
[Fe/H]\footnote{[A/B]$\equiv$log(N$_{\rm A}$/N$_{\rm B}$)$_{\rm star}$--log(N$_{\rm A}$/N$_{\rm B}$)$_{\sun}$ and log $\epsilon$(A)$\equiv$log(N$_{\rm A}$/N$_{\rm H}$)+12.0 for elements A and B.}, the various stellar populations are 
identified spectroscopically by their light element chemistry.  Specifically, 
stars within a single globular cluster are often categorized by their [O/Fe] 
and/or [Na/Fe] ratios as belonging to the ``primordial", ``intermediate", or 
``extreme" populations (e.g., Carretta et al. 2009).  In this categorization, 
primordial stars (first generation) are similar in composition to metal--poor 
halo field stars (O--rich; Na--poor), and intermediate and extreme stars 
(second generation) have lower [O/Fe] ratios and higher [Na/Fe] ratios.  The 
extreme stars are further distinguished from the intermediate population as 
having the lowest oxygen abundances ([O/Fe]$\la$--0.4) and highest sodium 
abundances ([Na/Fe]$\ga$$+$0.5).  While the intermediate population tends to 
dominate by number over the primordial population, the extreme stars are found 
only in a handful of globular clusters (Carretta et al. 2009; their Table 5).

The large and often (anti--)correlated abundance variations of elements 
ranging from carbon through aluminum is evidence that the material in
globular cluster stars' atmospheres has been subjected to high--temperature 
proton--capture nucleosynthesis (e.g., Denisenkov \& Denisenkova 1990; Langer 
et al. 1993; Prantzos et al. 2007).  While changes in [C/Fe], [N/Fe], and 
\iso{12}{C}/\iso{13}{C} ratios as a function of evolutionary state on the 
subgiant branch (SGB) and red giant branch (RGB) are clearly linked to 
\emph{in situ} mixing processes (e.g., Denissenkov \& VandenBerg 2003), the 
temperatures reached near the bottom of the convective envelope in more evolved
low--mass RGB stars are too low to significantly alter the abundances of heavier
elements (but see also D'Antona \& Ventura 2007 for a possible exception).  
Observations of similar star--to--star abundance variations among scarcely 
evolved globular cluster main--sequence and SGB stars (e.g., Briley et al.
1996; Gratton et al. 2001; Ram{\'{\i}}rez \& Cohen 2002; Ram{\'{\i}}rez \& 
Cohen 2003; Carretta et al. 2004; Briley et al. 2004; Cohen \& Mel{\'e}ndez 
2005; Bragaglia et al. 2010; D'Orazi et al. 2010; Dobrovolskas et al. 2014) 
indicate that the composition differences between the various globular cluster 
populations were already imprinted on the gas from which the stars formed.  
Although there is still no consensus regarding the nucleosynthesis source(s) 
driving the composition differences, possible options include intermediate 
mass ($\sim$5--8 M$_{\odot}$) asymptotic giant branch (AGB) stars (e.g., 
Karakas et al. 2006; Ventura \& D'Antona 2009), rapidly rotating massive 
($\ga$20 M$_{\odot}$) main--sequence stars (e.g., Decressin et al. 2007a), 
massive binary stars (de Mink et al. 2009), and super massive 
($\sim$10$^{\rm 4}$ M$_{\rm \odot}$) stars (Denissenkov \& Hartwick 2014).  
The physical process by which globular clusters form and evolve remains an 
open question as well (e.g., Decressin et al. 2007b; D'Ercole et al. 2008; 
Renzini 2008; Carretta et al. 2010a; Bekki 2011; Conroy \& Spergel 2011; 
Valcarce \& Catelan 2011; Bastian et al. 2013).

Regardless of the pollution source(s) in globular clusters, the nuclear
processes creating O--depleted and Na--enhanced gas may also be concurrent with
He enhancements.  With the exception of a few cases, such as NGC 1851, where 
the C$+$N$+$O sum may be variable (e.g., Ventura et al. 2009; Yong et al. 
2009; Yong et al. 2014), He abundance differences ranging from 
$\Delta$Y$\sim$0.02--0.20 have been invoked to explain many of the multiple 
photometric sequences observed in recent cluster color--magnitude diagrams 
(e.g., Piotto et al. 2007; Milone et al. 2012).  

Combined evidence from photometry and spectroscopy ties the traditional light 
element abundance patterns to varying levels of He enhancement (e.g., Bragaglia
et al. 2010; Dupree et al. 2011; Pasquini et al. 2011; Marino et al. 2014; 
Mucciarelli et al. 2014).  In addition to causing subtle effects in absorption 
line formation (B\"{o}hm--Vitense 1979), He enhancements can significantly 
alter a star's position on the color--magnitude diagram and shorten its 
evolutionary timescale, relative to a He--normal star.  There are some 
indications that the extent (or existence) of the AGB phase may be sensitive 
to a star's initial He abundance such that He--enhanced stars may evolve off 
the horizontal branch to become AGB--manqu{\'e} stars (e.g., Greggio \& Renzini
1990; Castellani et al. 2006; Gratton et al. 2010; Charbonnel et al. 2013).  In
particular, several authors have noted a peculiar feature in many clusters that
CN--strong and O--poor/Na--rich stars appear with a lower frequency on the AGB 
compared to the RGB (Mallia 1978; Norris et al. 1981; Suntzeff 1981; Smith \& 
Norris 1993; Pilachowski et al. 1996; Ivans et al. 1999; Sneden et al. 2000; 
Campbell et al. 2006; Campbell et al. 2010; Smolinski et al. 2011; Johnson \& 
Pilachowski 2012; Campbell et al. 2012; Campbell et al. 2013).  However, it is 
not yet clear that the lack of CN--strong, O--poor, and Na--rich stars on the 
AGB is a ubiquitous property of all globular clusters.  

Gratton et al. (2010) note that the AGB/RGB ratio in globular clusters is 
correlated with the minimum mass along the horizontal branch.  In other words, 
clusters with redder horizontal branches, and possibly also lower 
levels of He enhancement, tend to retain a larger fraction of stars between 
the RGB and AGB phases.  Therefore, in order to investigate this phenomenon 
further we have obtained spectra of RGB and AGB stars in the red horizontal 
branch, and relatively metal--rich ([Fe/H]$\approx$--0.7), globular cluster 47 
Tucanae (47 Tuc), and compare the [Na/Fe] distributions between stars in the 
two evolutionary states.  47 Tuc exhibits a relatively high AGB/RGB ratio 
($\sim$0.10; Gratton et al. 2010), and is suspected of having a predominantly 
CN--strong/Na--rich AGB population (Mallia 1978).  If confirmed, a dominant 
Na--rich AGB population in 47 Tuc would strongly contrast with the 
completely Na--poor AGB population in the more metal--poor blue horizontal
branch cluster NGC 6752 (Campbell et al. 2013).

\section{OBSERVATIONS, TARGET SELECTION, AND DATA REDUCTION}

\subsection{Observations and Instrument Description}

The RGB and AGB data sets for this project were taken on separate nights and
with different instruments.  The RGB data were obtained in 2011 November using 
the FLAMES--GIRAFFE instrument on the VLT--UT2 telescope at the European
Southern Observatory on Cerro Paranal.  Additional details regarding the 
observation, reduction, and analysis of these data can be found in Cordero
et al. (2014; their Section 2).  The new observations of 47 Tuc AGB stars
presented here were taken on 2014 June 1 with the Michigan/Magellan Fiber 
System (M2FS; Mateo et al. 2012) mounted on the Nasmyth--East port of the 
Magellan--Clay 6.5m telescope at Las Campanas Observatory.  

While further details about M2FS can be found in Mateo et al. (2012), we
summarize here basic information about the instrument and the specific setup
used for this project.  M2FS is a wide--field (29.3$\arcmin$) fiber--fed 
multiobject system with two sets of 128 fiber bundles (256 total) 
feeding identical ``red" and ``blue" spectrographs\footnote{We stress that 
while the two spectrographs are referred to as ``red" and ``blue", this
nomenclature is for identification purposes only and does not reflect any 
differences in wavelength nor optimization.}.  The 1.2$\arcsec$ 
fibers\footnote{Note that 1.2$\arcsec$ refers to the aperture size at the 
front of the fiber.} are mounted by hand into machine drilled plug plates, and 
can be placed with a minimum distance of $\sim$13$\arcsec$ between fibers.  
For the high resolution mode used here, the spectra are created by passing 
light through both an echelle grating and cross--dispersing prism.  Although 
observing in a single order allows for up to 256 fibers to be used 
simultaneously, the cross--disperser permits order stacking at the expense of 
using fewer fibers.  Several post--fiber spectrograph slit widths are available
that range from 180--45$\mu$m and provide a resolving power of 
R$\equiv$$\lambda$/$\Delta$$\lambda$=20,000--38,000, respectively (Mateo et 
al. 2012).

\begin{figure}
\epsscale{1.00}
\plotone{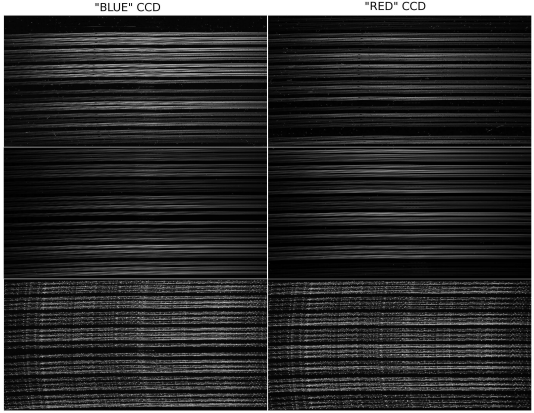}
\caption{Sample M2FS object (top), quartz flat (middle), and ThAr comparison
spectrum (bottom) images from the ``blue" (left) and ``red" (right) CCDs used
during the 47 Tuc observations.  Up to 24 fibers per CCD (48 total) can be used
with the M2FS setup employed here.  Each set of 6 orders corresponds to one
fiber.  For the image orientations shown here, wavelength increases from left
to right in each order.  H$\alpha$ can be seen in all of the stellar object
spectra.  Note also the internal Littrow ghost reflections affecting several
middle columns of each CCD.  This affects a roughly 5--10 \AA\ wide region in
the reduced spectra of some orders and some fibers.}
\label{f1}
\end{figure}

\begin{figure}
\epsscale{1.00}
\plotone{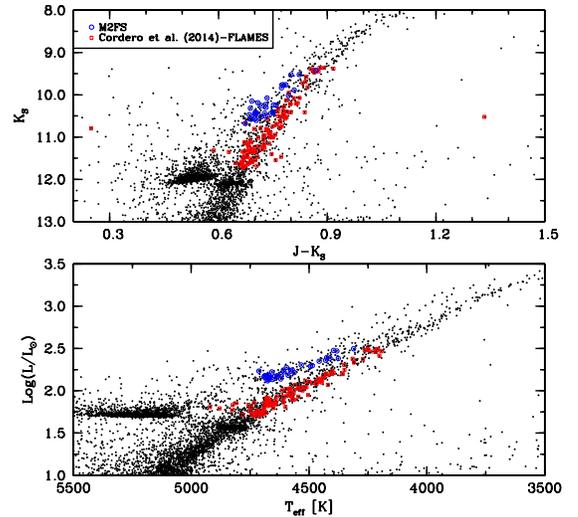}
\caption{\emph{top:} Color--magnitude diagram illustrating the evolutionary
state of AGB stars targeted in this work (open blue circles) and the
complimentary RGB FLAMES--GIRAFFE data set (open red boxes) from Cordero et al.
(2014).  The filled black circles are from 2MASS (Skrutskie et al. 2006).
\emph{bottom:} A temperature--luminosity plot showing the selection of RGB and
AGB stars from the McDonald et al. (2011) data.}
\label{f2}
\end{figure}

Specific to this project, we developed a spectrograph setup mode that employs a 
wide band order blocking filter (``Bulge$\_$GC1") providing continuous 
wavelength coverage from $\sim$6120--6720 \AA\ over 6 consecutive orders 
(58--53; see also Table 1).  This reduces the maximum number of available 
fibers from 256 to 48 (24 per spectrograph channel), but is roughly equivalent 
in terms of efficiency to a system such as FLAMES--GIRAFFE.  Sample object, 
quartz flat, and ThAr comparison spectrum images obtained for this project are 
shown in Figure \ref{f1}.  An internal Littrow ghost reflection (e.g.,
Burgh et al. 2007) affecting the middle columns of both 
CCDs was discovered when using this setup; however, the reflection only 
interferes with a roughly 5--10 \AA\ wide region in some orders of some fibers 
and is easily avoided in the reduced spectra.  Our observations used a 4 amp 
slow readout, 2$\times$1 (spatial$\times$dispersion) binning, and the 125$\mu$m
slits to achieve a resolving power of R=22,500.  We placed 42 fibers on 
potential 47 Tuc AGB stars and 5 fibers on blank sky regions.  Only one 
available fiber was left unassigned.  Since the AGB stars in 47 Tuc are 
relatively bright (V$\la$14; see also Figure \ref{f2}) and the observing 
conditions were good (FWHM$\la$1$\arcsec$), a set of 3$\times$1200 second 
exposures was sufficient to produce a signal--to--noise ratio (S/N) of about 
100 per resolution element.

\subsection{Target Selection}

The initial target selection and RGB/AGB separation was accomplished using
the photometry, luminosity, and temperature values from McDonald et al. (2011).
Since the data presented in Cordero et al. (2014) contained almost exclusively
RGB stars, we selected only early AGB stars of comparable temperature 
from the McDonald et al. (2011) catalog for observation with M2FS.  The 
data from McDonald et al. (2011) were matched to the Two Micron All--Sky
Survey (2MASS; Skrutskie et al. 2006) database, and the 2MASS coordinates were 
used as input for the fiber plug plate configuration.  Similarly, 2MASS 
coordinates and photometry were used to identify suitable guide, acquisition, 
and Shack--Hartmann stars required by the instrument.  Although we obtained 
spectra of 42 AGB stars, 7 were discarded due to a failure to adequately 
converge to a stable solution of the model atmosphere parameters (see Section 
3).  A 2MASS color--magnitude diagram and temperature--luminosity plot for the 
final sample of 35 AGB stars, along with the comparative sample of 113 RGB 
stars from Cordero et al. (2014), are shown in Figure \ref{f2}.  The star 
identifications, coordinates, photometry, model atmosphere parameters, 
abundances, and heliocentric radial velocities for all 35 AGB stars are 
provided in Table 2.

\begin{figure}
\epsscale{1.00}
\plotone{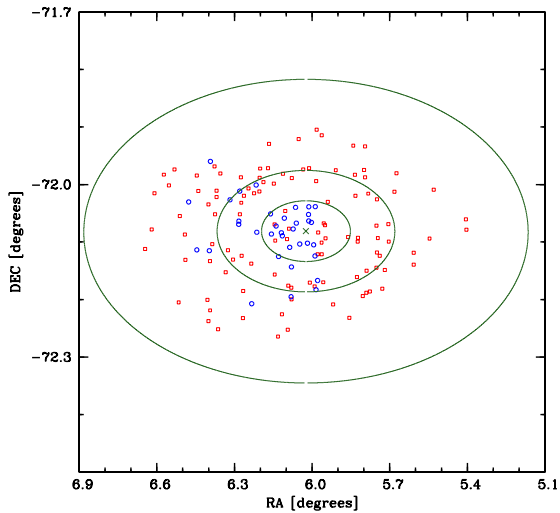}
\caption{Plot showing the sky coordinates of the AGB stars observed in this
work (open blue circles) and the RGB FLAMES--GIRAFFE data (open red boxes)
from Cordero et al. (2014).  The green cross indicates the cluster center, and
the outer ellipses designate 1, 2, and 5 times the half--light radius
(3.17$\arcmin$; Harris 1996).}
\label{f3}
\end{figure}

Figure \ref{f3} compares the sky positions of the AGB and RGB 47 Tuc samples
relative to the cluster center and half--light radius (r$_{\rm h}$).  We 
prioritized AGB stars residing within $\sim$1--2 r$_{\rm h}$, but also sampled 
out to the same radial extent as the RGB data.  The radial distribution of 
targets is relevant because one of the key science questions addressed in 
this paper is whether or not second generation RGB stars fail to evolve off
the horizontal branch and ascend the AGB.  There is growing evidence that
while primordial (first generation) stars tend to follow the underlying cluster
distribution at all radii, the intermediate and extreme stars (second 
generation) are often strongly concentrated near the cluster core (e.g., 
Carretta et al. 2010b; Kravtsov et al. 2010; Lardo et al. 2011; Johnson \& 
Pilachowski 2012; Milone et al. 2012; Cordero et al. 2014; Li et al. 2014).
Nataf et al. (2011) also find evidence that stars near the cluster core may
be He enhanced (see also di Criscienzo et al. 2010).  If He and Na are 
correlated then the most Na--rich AGB stars should reside near the core.
Vesperini et al. (2013) note that the local ratio of second/first generation 
stars should match the global ratio at a half--mass radius\footnote{The 
half--light and half--mass radii are roughly equivalent in 47 Tuc.} of 
$\sim$1.5.  Therefore, we expect that our AGB target selection will adequately 
sample the true AGB [Na/Fe] distribution and maximize our chances of finding 
Na--rich AGB stars, if they exist.

\subsection{Data Reduction}

The basic data reduction procedures, including bias correction and trimming the
overscan regions, were carried out using the IRAF\footnote{IRAF is distributed 
by the National Optical Astronomy Observatory, which is operated by the 
Association of Universities for Research in Astronomy, Inc., under cooperative 
agreement with the National Science Foundation.} task \emph{ccdproc}.  Since
each amplifier image on each CCD has a slightly different value for read noise
and gain, the individual amplifier object and calibration files are 
reduced separately in a manner similar to that used for Hectochelle 
reductions (Caldwell et al. 2009; Szentgyorgyi et al. 2011).  The individual 
bias corrected and trimmed exposures are then rotated, translated, and 
combined using the IRAF tasks \emph{imtranspose} and \emph{imjoin} to create 
one file per exposure (i.e., images such as those in Figure \ref{f1}).  

Although the M2FS setup used here produces 6 orders per fiber, the more 
advanced data reduction procedures should in principle be similar to the 
default setup in which each fiber produces a single order.  Therefore, we used
repeated calls of the IRAF task \emph{dohydra} to handle aperture 
identification and tracing, scattered light removal, flat--field correction,
ThAr wavelength calibration, cosmic--ray removal, and object spectrum 
extraction.  We ran \emph{dohydra} on all fibers of each CCD, but only 
extracted one order from each fiber per \emph{dohydra} loop.  Since the sky 
fibers are spread across both CCDs, we skipped the sky subtraction routine 
inside \emph{dohydra}.  Instead, we used \emph{scombine} to create a master sky
spectrum for each order of each exposure and subtracted this from the object 
spectra with the \emph{skysub} routine.  

\begin{figure}
\epsscale{0.80}
\plotone{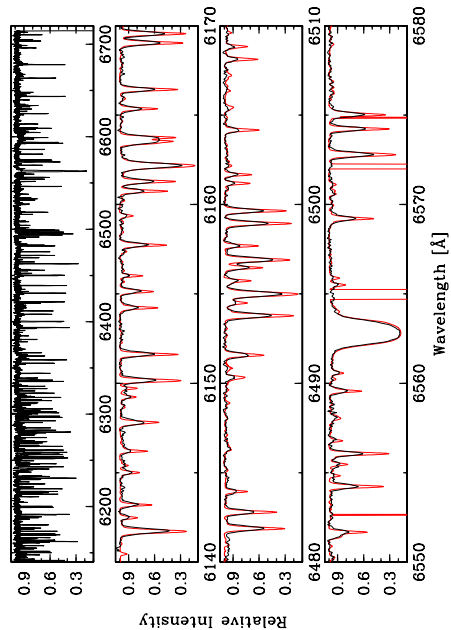}
\caption{Sample M2FS spectrum of a 47 Tuc AGB star after merging the orders.
The top panel shows the full wavelength range.  The lower panels show
sample spectral regions, including the 6154/6160 \AA\ Na I lines used to derive
the Na abundances for this work.  The higher resolution Arcturus atlas (Hinkle
et al. 2000) is shown in red for comparison.  Both stars have similar effective
temperatures.}
\label{f4}
\end{figure}

Following sky subtraction, each order of every exposure was normalized with 
the \emph{continuum} IRAF routine and then median combined with \emph{scombine}
to increase the S/N and remove any remaining cosmic--rays.  The full
wavelength range spanned by each order is listed in Table 1.  However, as 
is evident in Figure \ref{f1}, the S/N decreases near the edges of each order.
Therefore, we removed the lower S/N regions of each order that had 
correspondingly higher S/N regions in an adjacent order.  The final 
(``effective") wavelength coverage of each order after carrying out this 
procedure is given in Table 1.  We also chose to remove, rather than
combine, overlapping regions between orders because there is a small difference
in resolution between the bluer and redder orders.  A sample M2FS spectrum 
with all orders merged is shown in Figure \ref{f4}.

\section{DATA ANALYSIS}

\subsection{Model Stellar Atmospheres}

\begin{figure}
\epsscale{1.00}
\plotone{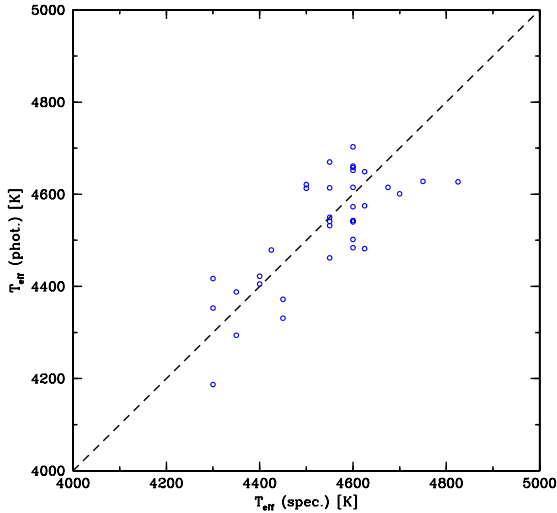}
\caption{Plot comparing the effective temperature values derived by enforcing
excitation equilibrium (T$_{\rm eff}$ spec.) and using the J--K$_{\rm S}$
color--temperature relation from Gonz{\'a}lez Hern{\'a}ndez \&
Bonifacio (2009; T$_{\rm eff}$ phot.).  The dashed line indicates perfect
agreement.}
\label{f5}
\end{figure}

The model atmosphere parameters effective temperature (T$_{\rm eff}$), surface
gravity (log(g)), metallicity ([Fe/H]), and microturbulence (vt) were 
determined spectroscopically and in a manner identical to that used for the 
FLAMES--GIRAFFE sample of Cordero et al. (2014).  Specifically, T$_{\rm eff}$ 
was determined by removing trends in log $\epsilon$(Fe I) as a function of 
excitation potential, log(g) was determined by enforcing ionization 
equilibrium with the Fe I and Fe II lines, and vt was determined by removing 
trends in log $\epsilon$(Fe I) as a function of line strength.  We adopted a 
generic model atmosphere with T$_{\rm eff}$=4600 K, log(g)=1.5 (cgs), 
[Fe/H]=--0.7, and vt=1.5 km s$^{\rm -1}$ as the starting point and then 
iterated to simultaneously solve all four model atmosphere parameters.  The 
average of the derived [Fe I/H] and [Fe II/H] abundances was used to update 
the model metallicity in each iteration.  Opacity and atmosphere stratification 
discrepancies due to the difference between [Fe/H] and [M/H] were roughly 
accounted for by using the $\alpha$--enhanced ATLAS9 model atmospheres from 
Castelli \& Kurucz (2004)\footnote{The model atmosphere grid
can be accessed at: http://wwwuser.oat.ts.astro.it/castelli/grids.html.}.
We interpolated within the available ATLAS9 grid to obtain the final model
atmospheres used in the abundance analysis.  Since the range in both 
temperatures and luminosities spanned by the RGB and AGB samples is relatively
small (see Figure \ref{f2}), we have not adopted any corrections to either 
the model atmosphere parameters or abundance ratios due to departures either
from plane parallel geometry or local thermodynamic equilibrium (LTE).  
However, Figure \ref{f5} shows that the spectroscopic temperatures are well 
correlated with temperatures estimates from (J--K$_{\rm S}$)$_{\rm o}$ 
2MASS photometry.  Using the color--temperature relation given in Gonz{\'a}lez 
Hern{\'a}ndez \& Bonifacio (2009), we find an average difference between the
spectroscopic and photometric temperatures of 14 K ($\sigma$=85 K).  We found
a similar result for the RGB FLAMES--GIRAFFE sample in Cordero et al. (2014),
with an average difference of 28 K ($\sigma$=53 K).

\subsection{Iron Abundance Determinations}

The line list used to determine Fe I and Fe II abundances was similar to that
used by Cordero et al. (2014), but augmented to include additional lines 
available in the M2FS spectra.  The [Fe/H] abundances, and 
also model atmosphere parameters, were based on equivalent width measurements 
for an average of 45 Fe I lines and 4 Fe II lines per star.  The equivalent 
widths were measured manually using the semi--automated code developed for 
Johnson et al. (2014), which fits individual or blended Gaussian profiles and 
is aided by a simple machine learning algorithm.  The final abundances were 
calculated using the \emph{abfind} driver of the LTE line analysis code 
MOOG\footnote{MOOG can be downloaded from 
http://www.as.utexas.edu/$\sim$chris/moog.html.} (Sneden 1973; 2010 version).  
We adopted the same solar abundance as Cordero et al. (2014) of log 
$\epsilon$(Fe)$_{\odot}$=7.52.  The list of lines and atomic data used for 
this work is provided in Table 3, and the log(gf) values were redetermined 
using a high S/N day light sky spectrum taken with M2FS.  For the lines in 
common between this work and Cordero et al. (2014), the average difference in 
adopted log(gf) values was zero for both Fe I ($\sigma$=0.09) and Fe II 
($\sigma$=0.04).

The random measurement uncertainty estimated by $\sigma$/$\sqrt{N}$, where 
$\sigma$ is the line--to--line dispersion and N is the number of lines 
measured, is relatively small for both iron species.  Specifically, the average
random measurement uncertainty for log $\epsilon$(Fe I) is 0.02 and for 
log $\epsilon$(Fe II) is 0.04.  The [Fe/H] error column listed in Table 2 
represents the combined uncertainty for both Fe I and Fe II abundances.
Additional uncertainty in the iron abundance determination comes from errors 
in the model atmosphere parameters.  We estimate that the uncertainty in 
T$_{\rm eff}$, log(g), [M/H], and vt, based solely on enforcing excitation 
equilibrium, ionization equilibrium, and removing trends in log $\epsilon$(Fe 
I) as a function of line strength, are: 50 K, 0.10 (cgs), 0.05 dex, and 0.10 km 
s$^{\rm -1}$, respectively.  The combined sensitivity of both [Fe I/H] and 
[Fe II/H] to these changes in model atmosphere parameters is 0.09 dex, on 
average.

As mentioned in Section 3.1, we did not apply any corrections to the [Fe I/H] 
and [Fe II/H] abundances for departures from LTE.  However, Lapenna et al.
(2014) used high resolution (R$\sim$48,000) spectra of 24 AGB and 11 RGB stars
in 47 Tuc to derive [Fe I/H] and [Fe II/H] abundances and found that only in
the RGB sample did the neutral and singly ionized abundances match.  Lapenna
et al. (2014) further conclude that for AGB stars Fe I lines should not be 
used to derive [Fe/H] abundances and that determining surface gravity from 
ionization equilibrium, which is the method used here, is also invalid for AGB 
stars.  The strong difference in Fe I line formation due to departures from LTE
in similar temperature AGB but not RGB stars is a puzzling result and does not 
match recent theoretical investigations (e.g., Bergemann et al. 2012; Lind et 
al. 2012).  We note that Ivans et al. (2001) also encountered problems
deriving model atmosphere parameters for evolved stars in the more metal--poor
([Fe/H]$\approx$--1.2) globular cluster M5, and also decided to set the [Fe/H]
scale using only Fe II and photometric log(g) values.

For the lines analyzed in this work, the Bergemann et al. (2012) and Lind et 
al. (2012) calculations suggest departures from LTE should affect [Fe I/H] at 
about the 0.05 dex level in both RGB and AGB stars\footnote{The non--LTE 
abundance corrections were calculated using the ``INSPECT" website interface 
(http://inspect-stars.net/).  Only 10 Fe I lines and 5 Fe II lines were 
available in both the line list used here and the INSPECT website.}.  In both 
populations the [Fe II/H] abundance is mostly unaffected because Fe II is the 
dominant ionization state.  However, the \emph{difference} in [Fe/H] 
abundances and model atmosphere parameters derived for RGB and AGB stars using
ionization and excitation equilibrium is not expected to produce large 
systematic offsets when assuming LTE.  In fact, after reconciling a small
difference in temperature scale between our RGB and AGB samples (see 
Section 4), we do not find any significant difference in [Fe/H] for the two
populations.  

Finally, we note that Lapenna et al. (2014) and this work have four AGB stars 
in common.  A comparison of the model atmosphere parameters, [Fe/H] abundances,
and radial velocities is provided in Table 4.  We find average differences in 
T$_{\rm eff}$, log(g), [Fe I/H], [Fe II/H], vt, and RV$_{\rm helio.}$ of 50 K, 
--0.12 (cgs), $+$0.12 dex, $+$0.05 dex, --0.09 km s$^{\rm -1}$, and --0.11 km 
s$^{\rm -1}$, respectively.  The surface gravity differences are even smaller 
if the most highly discrepant star (2M00235852--7206177) is removed.  The 
[Fe I/H] values from Lapenna et al. (2014) are the most discrepant abundances; 
however, we derive similar [Fe II/H] abundances (within the stated 
uncertainties) to Lapenna et al. (2014) while simultaneously solving for the 
model atmosphere parameters via spectroscopic methods and obtaining identical 
[Fe I/H] and [Fe II/H] ratios.  The conflicting observational claims between
this work and Lapenna et al. (2014) suggest that a better understanding of
RGB and AGB atmospheres is needed before ionization equilibrium is fully
dismissed as a viable method for determining [Fe/H] and surface gravity in AGB 
(but not RGB) stars.  However, we cannot rule out that the failure of 7 AGB 
stars in our sample (17$\%$) to converge to a stable spectroscopic solution is 
related to the discrepancies found by Lapenna et al. (2014) and Ivans et al. 
(2001).

\subsection{Sodium Abundance Determinations}

\begin{figure}
\epsscale{0.40}
\plotone{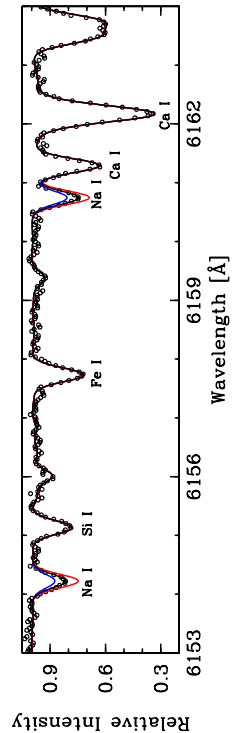}
\caption{Sample Na I spectrum synthesis for a typical 47 Tuc AGB star.  The
black line is the best fit synthesis and the red/blue lines are the best
fit Na abundance changed by $\pm$0.3 dex.}
\label{f6}
\end{figure}

For consistency with the Cordero et al. (2014) [Na/Fe] data, we measured 
sodium abundances using the same line list (see also Table 3 for a summary of
the transition parameters for the Na I lines), the \emph{synth} 
driver in MOOG, and adopted log $\epsilon$(Na)$_{\odot}$=6.33.  The dominant 
molecular contaminator near the 6154/6160 \AA\ Na I lines in the temperature 
and metallicity regime analyzed here is CN.  While we did not have individual 
C, N, and O abundances for the target stars, the local CN features were fit 
assuming that the stars were well--mixed (e.g., [C/Fe]=--0.5; [O/Fe]=$+$0.1; 
\iso{12}{C}/\iso{13}{C}=4); the nitrogen abundance was varied as a free 
parameter.  The CN line list was adopted from the Kurucz (1994) database (but 
see also a recent update by Sneden et al. 2014).  A sample spectrum synthesis 
for a typical M2FS spectrum is shown in Figure \ref{f6}.  The final [Na/Fe] 
values given in Table 2 do not include any corrections for departures from LTE 
in the analysis.  However, the log $\epsilon$(Na) non--LTE corrections are 
$\la$0.10 dex in an absolute sense (e.g., Lind et al. 2011; Thygesen et al. 
2014), and the differential non--LTE corrections for [Na/Fe] due to surface 
gravity differences between AGB and RGB stars is typically $<$0.05 dex.

The random measurement errors for log $\epsilon$(Na I) are only slightly larger
than for $\epsilon$(Fe I), with an average of 0.03 dex.  This value reflects
the difference in sodium abundance between the 6154 and 6160 \AA\ Na I lines.
The [Na/Fe] error column of Table 2 takes into account both the measurement
uncertainty in [Fe/H] and [Na/H].  As with iron, additional sources of 
uncertainty are due to errors in the derived model atmosphere parameters.
Using the same estimates as in Section 3.2, the average uncertainty from 
model atmosphere parameters alone in the [Na/Fe] ratio is 0.05 dex.

\section{BASIC RESULTS}

Although recent color--magnitude diagrams of 47 Tuc suggest the cluster has a 
complex star formation history (Anderson et al. 2009; di Criscienzo et al. 
2010; Nataf et al. 2011; Milone et al. 2012; Monelli et al. 2013; Li et al. 
2014), there is no evidence supporting a substantial spread in [Fe/H].  
Instead, literature work suggests the cluster has an average 
[Fe/H]$\approx$--0.60 to --0.80, with a star--to--star dispersion $\la$0.10 dex
(e.g., Carretta \& Gratton 1997; Alves--Brito et al. 2005; Wylie et al. 2006;
Koch \& McWilliam 2008; Carretta et al. 2009; Gratton et al. 2013; Cordero et 
al. 2014; Dobrovolskas et al. 2014; Thygesen et al. 2014).  In agreement with 
past work, we find an average [Fe/H]=--0.68 and a star--to--star dispersion of 
$\sigma$=0.08.  This result is similar to, but more metal--rich than, the RGB 
FLAMES--GIRAFFE sample of Cordero et al. (2014) that found 
$\langle$[Fe/H]$\rangle$=--0.75 ($\sigma$=0.10).  The 0.07 dex difference in 
[Fe/H] between the AGB and RGB samples is possibly tied to a systematic offset 
in the temperature scales between the two data sets.  A comparison of the 
spectroscopically determined T$_{\rm eff}$ values with those derived from 
photometric spectral energy distribution fitting in McDonald et al. (2011) 
finds systematic offsets of --20 K ($\sigma$=74 K) for the RGB sample and $+$35
K ($\sigma$=71 K) for the AGB sample.  If the [Fe/H] abundances are 
redetermined using temperatures (and corresponding gravities) that are 35 K 
lower and 20 K higher for the AGB and RGB stars, respectively, then the AGB 
[Fe/H] values decrease by 0.05 dex and the RGB [Fe/H] values increase by 0.02 
dex.  This brings both data sets into agreement at 
$\langle$[Fe/H]$\rangle$=--0.73.

In contrast to [Fe/H], 47 Tuc exhibits a significant spread in [Na/Fe] 
abundance.  For the AGB stars analyzed here, [Na/Fe] ranges from --0.11 to
$+$0.62 with $\langle$[Na/Fe]$\rangle$=$+$0.21 ($\sigma$=0.17).  The large
dispersion in [Na/Fe] is typical for globular clusters (e.g., see reviews by
Kraft 1994; Gratton et al. 2004; 2012), and the large fraction of stars with 
[Na/Fe]$>$0 matches previous observations of 47 Tuc RGB, horizontal branch, 
and main--sequence turn--off stars (e.g., Carretta et al. 2004; Alves--Brito 
et al. 2005; Koch \& McWilliam 2008; Carretta et al. 2009; D'Orazi et al.
2010; Worley \& Cottrell 2012; Gratton et al. 2013; Cordero et al. 2014; 
Dobrovolskas et al. 2014; Thygesen et al. 2014).  The possible temperature 
scale difference between the AGB and RGB samples also affects sodium.  However,
the change in [Na/Fe] is only an increase of 0.02 dex for the AGB stars and a 
decrease of 0.01 dex for the RGB stars.  A more detailed comparison between the
AGB and RGB [Na/Fe] abundances is provided in Section 5.

Radial velocities were also measured for each star in order to verify cluster
membership.  We used the cross--correlation code XCSAO (Kurtz \& Mink 1998), 
and the heliocentric correction was determined using the IRAF task 
\emph{rvcorrect}.  The velocities were determined relative
to a synthetic rest frame spectrum with T$_{\rm eff}$=4600 K, log(g)=1.5 (cgs),
[Fe/H]=--0.70, and vt=1.5 km s$^{\rm -1}$, and the XCSAO routine was only run 
on order 55 (see Table 1) of a single exposure for each star.  Order 55 was 
selected because it exhibits several strong lines but does not contain many
problematic features (e.g., telluric bands; unusually strong or broad lines).
The average measurement uncertainty returned by XCSAO for the radial velocities
was 0.18 km s$^{\rm -1}$.  We find an average heliocentric radial velocity of 
--18.56 km s$^{\rm -1}$ ($\sigma$=10.21 km s$^{\rm -1}$).  The average cluster 
velocity determined here is in good agreement with past work, which ranges 
from --16.86 km s$^{\rm -1}$ to --22.43 km s$^{\rm -1}$ (Dinescu et al. 1999; 
Carretta et al. 2004; Alves--Brito et al. 2005; Carretta et al. 2009; 
Dobrovolskas et al. 2014; Lapenna et al. 2014).  All of the stars observed 
with M2FS have velocities consistent with cluster membership.

\section{INTERPRETING THE AGB AND RGB [NA/FE] DISTRIBUTIONS}

Recent work mentioned in Section 1 suggests that AGB and RGB stars in some 
globular clusters may show distinctly different chemical compositions.
In particular, the AGB populations are suspected of having systematically
lower ratios of second generation (CN--strong; Na--enhanced; He--enhanced)
to first generation (CN--weak; Na--poor; He--normal) stars, compared to the 
RGB populations.  Evidence linking the minimum mass and blue extension of stars
along the horizontal branch to the AGB/RGB star count ratio (Gratton et al. 
2010) indicates that the apparent composition difference between AGB and RGB 
stars is likely due to some RGB stars not evolving through the AGB phase, 
rather than an \emph{in situ} process altering the envelope composition of 
some stars before or after the horizontal branch phase.  The apparent loss of 
Na--rich second generation stars between the RGB and AGB is most evident in the 
metal--poor blue horizontal branch cluster NGC 6752, which has a 
Na--poor/Na--rich ratio of 30:70 on the RGB and a 100:0 ratio on the AGB 
(Campbell et al. 2013).  Furthermore, Sandquist \& Bolte (2004) note that NGC 
6752 exhibits a low ratio of AGB to horizontal branch stars, and interpret 
this observation as an indication that some fraction of RGB stars fail to 
evolve through the AGB phase.

\begin{figure}
\epsscale{1.00}
\plotone{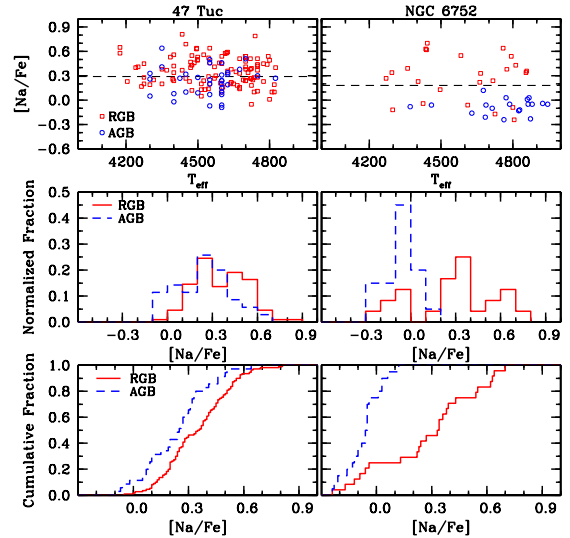}
\caption{Plots of [Na/Fe] versus T$_{\rm eff}$ (top), [Na/Fe] histograms using
0.1 dex bins (middle), and cumulative distribution functions (bottom) are
shown for 47 Tuc (left) and NGC 6752 (right).  While the 47 Tuc AGB data
are from this work, the 47 Tuc RGB data are from the FLAMES--GIRAFFE sample
in Cordero et al. (2014) and the RGB/AGB data for NGC 6752 are from
Campbell et al. (2013).  The dashed horizontal lines in the top panels
separate the Na--poor and Na--rich populations, based on the inflection point
in the cumulative distribution panels.  Typical measurement errors for [Na/Fe]
are $\sim$0.1 dex.  Note that for 47 Tuc the AGB [Na/Fe] abundances have been
increased by 0.02 dex and the RGB abundances, including the Na--poor/Na--rich
separation, decreased by 0.01 dex, in order to bring the two samples onto a
common temperature scale (see also Section 4).}
\label{f7}
\end{figure}

Since the number of dedicated studies comparing RGB and AGB abundance patterns
is still small (e.g., see Campbell et al. 2006; their Table 1), evidence is 
insufficient to determine whether the peculiar lack of Na--rich AGB 
stars in NGC 6752 is more the exception or the rule.  Therefore, the data 
provided here offer insight into a possible counter example.  47 Tuc is 
approximately 10$\times$ more massive than NGC 6752 (e.g., Pryor \& Meylan 
1993), 7$\times$ more metal--rich, contains almost exclusively red horizontal 
branch stars (e.g., Lee et al. 1994; see also Figure \ref{f2}), and has a high 
AGB/RGB ratio (Gratton et al. 2010).  In the left panels of Figure \ref{f7}, we
compare the [Na/Fe] abundances between the AGB and RGB samples in 47 Tuc as a 
function of T$_{\rm eff}$, as a binned distribution, and as a cumulative 
distribution.  The AGB [Na/Fe] distribution appears to be shifted to lower
values than the RGB population, and also has fewer stars with [Na/Fe]$>$$+$0.4
but more with [Na/Fe]$<$$+$0.1.  After correcting for the AGB/RGB temperature
scale differences, comparing only the average values yields
$\langle$[Na/Fe]$\rangle$=$+$0.23 for the AGB sample and 
$\langle$[Na/Fe]$\rangle$=$+$0.35 for the RGB FLAMES--GIRAFFE sample.  
Similarly, the ratio of Na--poor to Na--rich stars is 63:37 for the AGB sample 
and 45:55 for the RGB sample\footnote{The division of Na--poor and Na--rich 
stars in 47 Tuc is set at [Na/Fe]=$+$0.29, after accounting for the AGB/RGB
temperature scale difference, and is based on the RGB [Na/Fe] distribution 
(see also Figure \ref{f7}).}.  The 0.12 dex average [Na/Fe] abundance 
difference between AGB and RGB stars in 47 Tuc is similar to, though less 
extreme than, the 0.34 dex average difference found by Campbell et al. (2013) 
for NGC 6752 (see also the right panels of Figure \ref{f7}).

However, an examination of the dispersion ($\sigma$) and interquartile range 
(IQR) values of the AGB and RGB stars in 47 Tuc, which are mostly insensitive 
to systematic reduction or analysis differences between data sets, reveals that
the two populations are similar.  Specifically, the 47 Tuc AGB stars have 
$\sigma$$_{\rm AGB}$=0.17 and IQR$_{\rm AGB}$=0.23 and the RGB stars have 
$\sigma$$_{\rm RGB}$=0.18 and IQR$_{\rm RGB}$=0.27.  These values strongly
contrast with the results of NGC 6752 in which the AGB has 
$\sigma$$_{\rm AGB}$=0.09 and IQR$_{\rm AGB}$=0.10 and the RGB has 
$\sigma$$_{\rm RGB}$=0.28 and IQR$_{\rm RGB}$=0.37.  The stark contrast in 
AGB/RGB [Na/Fe] dispersion and IQR distributions between the two clusters 
suggests that only a small fraction ($<$20$\%$) of Na--rich RGB stars in 47 Tuc
fail to evolve through at least the early AGB phase, instead of the 100$\%$ 
fraction observed in NGC 6752.

\begin{figure}
\epsscale{1.00}
\plotone{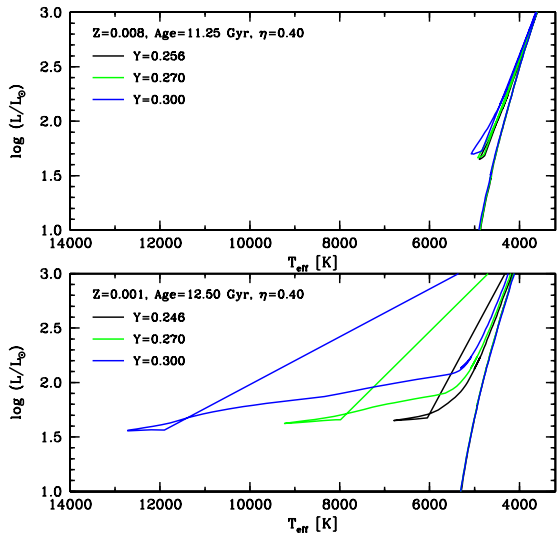}
\caption{\emph{top:} Stellar isochrones from Bertelli et al. (2008) are shown
assuming the age and metallicity of 47 Tuc from VandenBerg et al. (2013).  The
black (Y=0.256), green (Y=0.270), and blue (Y=0.300) isochrones illustrate how
the horizontal branch changes as a function of He abundance.  A Reimers (1975)
mass loss constant ($\eta$) of 0.4 is assumed for all cases.  \emph{bottom:} A
similar plot assuming the age and metallicity of NGC 6752.}
\label{f8}
\end{figure}

Color--magnitude diagram analyses also support the notion that both
Na--poor and Na--rich RGB stars evolve through the horizontal branch and 
early AGB phases in 47 Tuc.  Aside from previous spectroscopic detections of 
both CN--strong (Na--rich) and CN--weak (Na--poor) stars on 47 Tuc's AGB (e.g., 
Mallia 1978; Briley 1997), we note that Monelli et al. (2013) make use of the 
(U--B)--(B--I) color index, which is correlated with [Na/Fe], and find a 
bimodal AGB color spread.  These data indicate that at least two stellar 
populations, containing roughly equal proportions but with different light 
element chemistry, are present along 47 Tuc's AGB.  Furthermore, 
color--magnitude diagrams do not show a significant population of blue 
horizontal branch stars that may fail to ascend the AGB (e.g., Anderson et al. 
2009; Bergbusch \& Stetson 2009; Dieball et al. 2009; Milone et al. 2012),
and stars spanning the entire RGB [Na/Fe] range are found on the red 
horizontal branch (Gratton et al. 2013).  This contrasts with the 
color--magnitude diagram of NGC 6752, which shows a significant population of 
blue and extreme horizontal branch stars (e.g., Grundahl et al. 1999), and also
is known to contain AGB--manqu{\'e} stars (e.g., Momany et al. 2002).  However,
the horizontal branch differences between 47 Tuc and NGC 6752 are largely 
driven by metallicity, as is qualitatively illustrated in the isochrone tracks 
of Figure \ref{f8}.

Since it is clear from Figure \ref{f7} that the RGB and AGB [Na/Fe] 
distributions are significantly different in absolute abundance \emph{and}
dispersion between 47 Tuc and NGC 6752, the question remains as to what ties 
together RGB composition and post--RGB evolution.  As mentioned previously, a 
commonly adopted solution is that Na--rich stars have a higher He abundance, 
and that the lower masses of He--rich/Na--rich horizontal branch stars prevent
them from evolving through the AGB phase.  At least for the extreme case of 
NGC 6752, Charbonnel et al. (2013) note that stars with masses less than about 
0.735 M$_{\odot}$ and Y$\ga$0.31 do not ascend the AGB.  However, the required 
He--enhancements are significantly larger than those estimated from photometry 
($\Delta$Y$\sim$0.03--0.04; Milone et al. 2013), but Charbonnel et al. (2013) 
further suggest the maximum photometric He spread could be underestimated.  
For 47 Tuc, the He spread is estimated to be even smaller 
($\Delta$Y$\la$0.02--0.03; e.g., di Criscienzo et al. 2010; Nataf et al. 2011;
Milone et al. 2012) than in NGC 6752, and as Figure \ref{f8} shows even RGB 
stars with Y=0.3 are still likely to ascend the AGB at 47 Tuc's metallicity
(e.g., see also Valcarce et al. 2012 and references therein).  Despite having 
significantly different metallicities and horizontal branch morphologies, the 
AGB/RGB composition difference in 47 Tuc may be more analagous to the case of 
M13 where Johnson \& Pilachowski (2012) found that only the most Na--rich 
\emph{and} O--poor stars (presumably also the most He--rich stars) failed to 
reach the AGB\footnote{Unfortunately, we were not able to measure [O/Fe] 
abundances for many stars in the AGB data set because the preferred 6300 \AA\ 
[O I] line was severely contaminated by telluric bands and the 6300 \AA\
sky emission feature.}.  We note that the paucity of O--poor stars on M13's 
AGB, albeit from a sample of only $\sim$10 AGB stars, is observed despite the 
failure (so far) of investigators to find a strongly He--enhanced population 
(Sandquist et al. 2010; Smith et al. 2014).  Could a different parameter be 
driving the AGB/RGB composition disparity?

Campbell et al. (2013) proposed that, in addition to He enhancement for the 
Na--rich stars, a substantial \emph{ad hoc} increase in the mass loss rate 
(20$\times$ the RGB value) for NGC 6752 stars on the blue horizontal branch 
may be required to prevent further evolution up the AGB.  However, Cassisi et 
al. (2014) note that the required mass loss rate of $\sim$10$^{\rm -9}$ 
M$_{\rm \odot}$ year$^{\rm -1}$ is several orders of magnitude higher than 
those allowed by current analyses of hot horizontal branch and B subdwarf 
stars.  Additionally, horizontal branch simulations by Cassisi et al. (2014)
that accurately reproduce the R$_{\rm 2}$ parameter (N$_{AGB}$/N$_{HB}$) for
NGC 6752, M3, and M13, but do not invoke enhanced mass loss, also predict that 
$\sim$50$\%$ of NGC 6752's AGB stars should be Na--rich (albeit with lower Y 
than the most extreme values).  These results indicate that enhanced mass loss 
for a particular subset of horizontal branch stars is unlikely to be the 
key link between RGB composition and post--RGB evolution.  Furthermore, even
if processes associated with the radiative levitation of metals in blue 
horizontal branch stars played a role in mass loss and/or the subsequent 
envelope composition of AGB stars, 47 Tuc lacks a significant population of 
horizontal branch stars with T$_{\rm eff}$$\ga$11,000 K that would experience
this effect (e.g., Behr et al. 2003).

\section{CONCLUSIONS}

The data presented here indicate that the average [Na/Fe] abundance of AGB 
stars in the Galactic globular cluster 47 Tuc is 0.12 dex lower than that of
the RGB stars.  Additionally, the ratio of Na--poor to Na--rich AGB stars is 
63:37 compared to 45:55 for the RGB stars.  However, both populations exhibit 
a similar dispersion and IQR of [Na/Fe].  This result strongly contrasts with 
that found by Campbell et al. (2013) in the globular cluster NGC 6752 where the 
AGB consists of only Na--poor stars with a small star--to--star [Na/Fe]
dispersion.  The 47 Tuc [Na/Fe] distribution suggests that $\la$20$\%$ 
of Na--rich RGB stars fail to reach the early AGB, which is similar to the 
case of M13 where only the most Na--rich and O--poor stars are missing from
the AGB.  For 47 Tuc, the cluster's relatively high metallicity leads to 
a predominantly red horizontal branch morphology, and no significant population
of hot horizontal branch and AGB--manqu{\'e} stars, which likely make up the 
missing Na--rich component of NGC 6752's (and M13's) AGB, has been found.  
Unlike the case for NGC 6752, however, in 47 Tuc it seems that at least some 
Na--rich RGB stars evolve through the early AGB phase.  We conclude that RGB 
[Na/Fe] abundances alone are not a unique predictor of future AGB evolution in 
all clusters.

Several remaining questions must be addressed before a definitive link between 
stellar composition, cluster horizontal branch morphology, and AGB evolution 
can be formed.  For example, if He abundance, in addition to metallicity and
age, is a significant factor in defining a globular cluster's horizontal branch 
morphology and a star's post--RGB evolution, why do the $\Delta$Y values 
estimated from photometry appear too low in a cluster such as NGC 6752 to 
prevent Na--rich stars from ascending the AGB?  Are the He--enhancements
actually underestimated, as was suggested by Charbonnel et al. (2013)?  What 
roles do additional parameters such as mass loss, He core rotation, CNO 
abundance, and/or environment play in defining a star's horizontal branch and 
eventual AGB evolution?  What separates clusters such as 47 Tuc and M13, which 
lose only a fraction of Na--rich stars before the AGB, to those such as NGC 
6752, which lose 100$\%$?  A deeper understanding of the critical parameters 
controlling horizontal branch and AGB evolution, along with additional studies 
comparing large samples of RGB and AGB chemical compositions, seems required in
order to fully place the failure (or not) of some Na--rich stars to ascend the 
AGB into context with the developing narrative of forming multiple populations 
in globular clusters.

\acknowledgements

We thank the anonymous referee for a thoughtful and constructive report that
led to improvement of the manuscript.  This research has made use of NASA's 
Astrophysics Data System Bibliographic Services.  This publication has made 
use of data products from the Two Micron All Sky Survey, which is a joint 
project of the University of Massachusetts and the Infrared Processing and 
Analysis Center/California Institute of Technology, funded by the National 
Aeronautics and Space Administration and the National Science Foundation.  
C.I.J. gratefully acknowledges support from the Clay Fellowship, administered 
by the Smithsonian Astrophysical Observatory.  C.I.J. would like to thank 
Andrea Dupree and Nelson Caldwell for helpful discussions regarding data 
reduction.  C.A.P. gratefully acknowledges support from the Daniel Kirkwood 
Research Fund at Indiana University.  M.C. is supported by 
Sonderforschungs--bereich SFB 881 ``The Milky Way System" (subproject A5) of 
the German Research Foundation (DFG).  M.M. is grateful for support from the 
National Science Foundation to develop M2FS (AST--0923160) and carry out the 
observations reported here (AST--1312997) and to the University of Michigan 
for its direct support of M2FS construction and operation.


\clearpage
\LongTables
\begin{landscape}

\tablenum{1}
\tablecolumns{3}
\tablewidth{0pt}

\begin{deluxetable}{ccc}
\tablecaption{M2FS Bulge\_GC1 Filter Orders}
\tablehead{
\colhead{Order} &
\colhead{Full Range}      &
\colhead{Effective Range}       \\
\colhead{}      &
\colhead{(\AA)}      &
\colhead{(\AA)}
}

\startdata
58      &       6122$-$6205     &       6138$-$6190     \\
57      &       6152$-$6315     &       6190$-$6310     \\
56      &       6262$-$6426     &       6310$-$6410     \\
55      &       6376$-$6543     &       6410$-$6525     \\
54      &       6494$-$6664     &       6525$-$6650     \\
53      &       6617$-$6721     &       6650$-$6720     \\
\enddata

\tablecomments{Note that orders 58 and 53 are only partially covered due to
the filter response cut--off.}

\end{deluxetable}
\clearpage

\tablenum{2}
\tablecolumns{15}
\tablewidth{0pt}

\begin{deluxetable}{ccccccccccccccc}
\tabletypesize{\scriptsize}
\tablecaption{Basic Data and Results}
\tablehead{
\colhead{Star ID}       &
\colhead{RA (J2000)}    &
\colhead{DEC (J2000)}      &
\colhead{J}      &
\colhead{H}      &
\colhead{K$_{\rm S}$}      &
\colhead{T$_{\rm eff}$}      &
\colhead{log(g)}      &
\colhead{[Fe/H]}        &
\colhead{[Fe/H] Error}  &
\colhead{vt}      &
\colhead{[Na/Fe]}        &
\colhead{[Na/Fe] Error}  &
\colhead{RV$_{\rm helio.}$}     &
\colhead{RV$_{\rm helio.}$ Error}       \\
\colhead{2MASS} &
\colhead{(degrees)}      &
\colhead{(degrees)}      &
\colhead{(mag.)}      &
\colhead{(mag.)}      &
\colhead{(mag.)}      &
\colhead{(K)}      &
\colhead{(cgs)}      &
\colhead{}      &
\colhead{}      &
\colhead{(km s$^{\rm -1}$)}      &
\colhead{}      &
\colhead{}      &
\colhead{(km s$^{\rm -1}$)}      &
\colhead{(km s$^{\rm -1}$)}
}

\startdata
2M00235484$-$7210009    &       5.978539        &       $-$72.166939    &       11.266  &       10.697  &       10.575  &       4600    &       1.71    &       $-$0.72 &       0.05    &       1.70    &       $+$0.18 &       0.05    &       $-$5.47 &       0.14    \\
2M00235632$-$7210590    &       5.984693        &       $-$72.183060    &       11.114  &       10.514  &       10.412  &       4500    &       1.60    &       $-$0.71 &       0.06    &       1.70    &       $+$0.26 &       0.08    &       $-$19.66        &       0.18    \\
2M00235701$-$7202185    &       5.987543        &       $-$72.038475    &       11.145  &       10.593  &       10.445  &       4825    &       2.18    &       $-$0.57 &       0.04    &       1.95    &       $+$0.25 &       0.05    &       $-$6.03 &       0.19    \\
2M00235728$-$7207274    &       5.988676        &       $-$72.124290    &       10.610  &       9.948   &       9.836   &       4400    &       1.47    &       $-$0.69 &       0.03    &       1.65    &       $+$0.06 &       0.07    &       $-$3.12 &       0.14    \\
2M00235852$-$7206177    &       5.993859        &       $-$72.104919    &       11.005  &       10.436  &       10.317  &       4600    &       1.33    &       $-$0.78 &       0.02    &       1.75    &       $+$0.44 &       0.03    &       $-$1.03 &       0.14    \\
2M00240062$-$7203573    &       6.002621        &       $-$72.065926    &       10.343  &       9.625   &       9.520   &       4350    &       1.17    &       $-$0.76 &       0.03    &       1.75    &       $+$0.39 &       0.08    &       $-$26.60        &       0.18    \\
2M00240304$-$7202193    &       6.012702        &       $-$72.038712    &       10.840  &       10.284  &       10.148  &       4625    &       1.93    &       $-$0.72 &       0.02    &       1.75    &       $+$0.17 &       0.06    &       $-$28.23        &       0.20    \\
2M00240310$-$7203482    &       6.012947        &       $-$72.063393    &       11.139  &       10.578  &       10.435  &       4600    &       1.92    &       $-$0.66 &       0.07    &       1.60    &       $+$0.08 &       0.07    &       $-$9.69 &       0.15    \\
2M00240330$-$7203075    &       6.013772        &       $-$72.052101    &       11.139  &       10.578  &       10.435  &       4550    &       1.67    &       $-$0.76 &       0.06    &       1.95    &       $+$0.48 &       0.06    &       $-$12.31        &       0.15    \\
2M00240427$-$7206074    &       6.017792        &       $-$72.102081    &       10.814  &       10.124  &       10.021  &       4450    &       1.73    &       $-$0.68 &       0.06    &       1.40    &       $+$0.08 &       0.06    &       $-$24.41        &       0.16    \\
2M00241142$-$7206126    &       6.047624        &       $-$72.103516    &       11.184  &       10.559  &       10.440  &       4600    &       1.72    &       $-$0.72 &       0.03    &       1.55    &       $+$0.05 &       0.03    &       $-$20.35        &       0.14    \\
2M00241462$-$7204018    &       6.060919        &       $-$72.067184    &       11.161  &       10.568  &       10.410  &       4600    &       1.85    &       $-$0.65 &       0.05    &       1.60    &       $+$0.22 &       0.06    &       $-$23.14        &       0.18    \\
2M00241531$-$7202231    &       6.063815        &       $-$72.039757    &       10.552  &       9.886   &       9.771   &       4400    &       1.42    &       $-$0.58 &       0.03    &       1.60    &       $-$0.04 &       0.04    &       $-$13.49        &       0.13    \\
2M00241755$-$7204385    &       6.073161        &       $-$72.077370    &       11.351  &       10.733  &       10.677  &       4600    &       2.11    &       $-$0.59 &       0.05    &       0.90    &       $-$0.08 &       0.06    &       $-$33.09        &       0.19    \\
2M00241912$-$7208360    &       6.079687        &       $-$72.143356    &       10.992  &       10.395  &       10.240  &       4425    &       1.13    &       $-$0.82 &       0.06    &       1.55    &       $+$0.25 &       0.06    &       $-$31.87        &       0.17    \\
2M00241945$-$7211426    &       6.081072        &       $-$72.195183    &       11.262  &       10.646  &       10.562  &       4750    &       2.03    &       $-$0.56 &       0.05    &       1.70    &       $+$0.28 &       0.05    &       $-$17.19        &       0.14    \\
2M00242074$-$7206332    &       6.086458        &       $-$72.109245    &       11.157  &       10.615  &       10.472  &       4550    &       1.27    &       $-$0.75 &       0.03    &       1.45    &       $+$0.05 &       0.03    &       $-$22.51        &       0.15    \\
2M00242572$-$7203307    &       6.107181        &       $-$72.058540    &       10.696  &       10.017  &       9.887   &       4450    &       1.15    &       $-$0.68 &       0.05    &       1.70    &       $+$0.30 &       0.05    &       $-$11.54        &       0.15    \\
2M00242763$-$7205213    &       6.115141        &       $-$72.089264    &       11.004  &       10.428  &       10.245  &       4550    &       1.64    &       $-$0.67 &       0.08    &       1.70    &       $+$0.18 &       0.09    &       $-$20.96        &       0.16    \\
2M00242846$-$7205019    &       6.118608        &       $-$72.083862    &       11.176  &       10.581  &       10.425  &       4625    &       1.65    &       $-$0.76 &       0.03    &       1.75    &       $+$0.32 &       0.05    &       $-$22.21        &       0.17    \\
2M00243106$-$7207311    &       6.129429        &       $-$72.125313    &       10.898  &       10.280  &       10.189  &       4700    &       1.90    &       $-$0.61 &       0.04    &       1.70    &       $+$0.49 &       0.05    &       $-$24.35        &       0.15    \\
2M00243337$-$7204200    &       6.139076        &       $-$72.072243    &       10.973  &       10.362  &       10.254  &       4600    &       1.95    &       $-$0.61 &       0.06    &       1.60    &       $+$0.29 &       0.07    &       $+$2.90 &       0.13    \\
2M00243771$-$7205100    &       6.157149        &       $-$72.086121    &       10.324  &       9.651   &       9.524   &       4300    &       1.20    &       $-$0.71 &       0.10    &       1.75    &       $+$0.31 &       0.10    &       $-$38.05        &       0.22    \\
2M00243851$-$7203038    &       6.160488        &       $-$72.051079    &       10.958  &       10.304  &       10.228  &       4550    &       1.48    &       $-$0.70 &       0.03    &       1.65    &       $+$0.23 &       0.04    &       $-$22.80        &       0.17    \\
2M00245127$-$7204588    &       6.213660        &       $-$72.083000    &       11.161  &       10.543  &       10.434  &       4550    &       1.50    &       $-$0.68 &       0.05    &       1.60    &       $+$0.44 &       0.06    &       $-$19.99        &       0.17    \\
2M00245189$-$7200031    &       6.216213        &       $-$72.000862    &       11.237  &       10.639  &       10.519  &       4625    &       1.63    &       $-$0.70 &       0.04    &       1.65    &       $+$0.30 &       0.06    &       $-$30.43        &       0.23    \\
2M00245600$-$7212266    &       6.233351        &       $-$72.207397    &       11.125  &       10.510  &       10.421  &       4675    &       2.15    &       $-$0.64 &       0.06    &       1.40    &       $+$0.36 &       0.07    &       $+$0.69 &       0.21    \\
2M00250716$-$7200415    &       6.279846        &       $-$72.011536    &       10.537  &       9.865   &       9.761   &       4300    &       1.37    &       $-$0.66 &       0.03    &       1.65    &       $+$0.03 &       0.03    &       $-$17.09        &       0.13    \\
2M00250791$-$7203490    &       6.282998        &       $-$72.063629    &       11.264  &       10.658  &       10.575  &       4600    &       1.52    &       $-$0.76 &       0.06    &       1.70    &       $+$0.13 &       0.07    &       $-$33.36        &       0.16    \\
2M00250809$-$7204092    &       6.283737        &       $-$72.069244    &       11.027  &       10.408  &       10.296  &       4600    &       1.69    &       $-$0.59 &       0.06    &       1.65    &       $+$0.22 &       0.06    &       $-$23.45        &       0.17    \\
2M00251614$-$7201359    &       6.317278        &       $-$72.026649    &       10.809  &       10.187  &       10.075  &       4550    &       1.43    &       $-$0.56 &       0.05    &       1.75    &       $-$0.09 &       0.11    &       $-$12.04        &       0.13    \\
2M00253428$-$7157352    &       6.392834        &       $-$71.959793    &       10.549  &       9.896   &       9.763   &       4350    &       1.08    &       $-$0.79 &       0.03    &       1.75    &       $+$0.62 &       0.04    &       $-$18.29        &       0.15    \\
2M00253529$-$7206543    &       6.397057        &       $-$72.115105    &       11.308  &       10.727  &       10.604  &       4500    &       1.38    &       $-$0.84 &       0.06    &       1.40    &       $+$0.07 &       0.06    &       $-$17.45        &       0.19    \\
2M00254689$-$7206494    &       6.445395        &       $-$72.113731    &       11.238  &       10.646  &       10.508  &       4600    &       2.01    &       $-$0.57 &       0.08    &       1.70    &       $-$0.11 &       0.08    &       $-$13.47        &       0.18    \\
2M00255446$-$7201491    &       6.476949        &       $-$72.030312    &       10.289  &       9.558   &       9.420   &       4300    &       1.27    &       $-$0.55 &       0.08    &       1.75    &       $+$0.24 &       0.12    &       $-$29.39        &       0.57    \\
\enddata

\end{deluxetable}
\clearpage

\tablenum{3}
\tablecolumns{4}
\tablewidth{0pt}

\begin{deluxetable}{cccc}
\tablecaption{Line List}
\tablehead{
\colhead{Wavelength}    &
\colhead{Ion}      &
\colhead{E.P.}  &
\colhead{log(gf)}       \\
\colhead{(\AA)}      &
\colhead{}      &
\colhead{(eV)}  &
\colhead{}
}

\startdata
6154.23 &       11.0    &       2.101   &       $-$1.57 \\
6160.75 &       11.0    &       2.103   &       $-$1.27 \\
6145.41 &       26.0    &       3.368   &       $-$3.58 \\
6148.65 &       26.0    &       4.320   &       $-$2.63 \\
6151.62 &       26.0    &       2.176   &       $-$3.31 \\
6157.73 &       26.0    &       4.076   &       $-$1.19 \\
6159.37 &       26.0    &       4.607   &       $-$1.88 \\
6165.36 &       26.0    &       4.143   &       $-$1.54 \\
6173.33 &       26.0    &       2.223   &       $-$2.89 \\
6180.20 &       26.0    &       2.727   &       $-$2.71 \\
6187.99 &       26.0    &       3.943   &       $-$1.75 \\
6213.43 &       26.0    &       2.223   &       $-$2.52 \\
6219.28 &       26.0    &       2.198   &       $-$2.33 \\
6220.78 &       26.0    &       3.881   &       $-$2.39 \\
6226.73 &       26.0    &       3.883   &       $-$2.12 \\
6229.23 &       26.0    &       2.845   &       $-$2.93 \\
6232.64 &       26.0    &       3.654   &       $-$1.24 \\
6240.65 &       26.0    &       2.223   &       $-$3.32 \\
6252.56 &       26.0    &       2.404   &       $-$1.63 \\
6253.83 &       26.0    &       4.733   &       $-$1.53 \\
6270.22 &       26.0    &       2.858   &       $-$2.62 \\
6271.28 &       26.0    &       3.332   &       $-$2.76 \\
6322.69 &       26.0    &       2.588   &       $-$2.31 \\
6335.33 &       26.0    &       2.198   &       $-$2.11 \\
6336.82 &       26.0    &       3.686   &       $-$0.59 \\
6358.70 &       26.0    &       0.859   &       $-$4.32 \\
6362.88 &       26.0    &       4.186   &       $-$2.03 \\
6380.74 &       26.0    &       4.186   &       $-$1.35 \\
6385.72 &       26.0    &       4.733   &       $-$1.87 \\
6392.54 &       26.0    &       2.279   &       $-$3.95 \\
6393.60 &       26.0    &       2.433   &       $-$1.48 \\
6408.02 &       26.0    &       3.686   &       $-$0.91 \\
6411.65 &       26.0    &       3.654   &       $-$0.40 \\
6430.85 &       26.0    &       2.176   &       $-$1.89 \\
6469.19 &       26.0    &       4.835   &       $-$0.80 \\
6472.15 &       26.0    &       4.371   &       $-$2.89 \\
6475.62 &       26.0    &       2.559   &       $-$2.88 \\
6481.87 &       26.0    &       2.279   &       $-$2.92 \\
6483.94 &       26.0    &       1.485   &       $-$5.30 \\
6494.98 &       26.0    &       2.404   &       $-$1.15 \\
6495.74 &       26.0    &       4.835   &       $-$1.02 \\
6496.47 &       26.0    &       4.795   &       $-$0.56 \\
6498.94 &       26.0    &       0.958   &       $-$4.61 \\
6509.62 &       26.0    &       4.076   &       $-$2.88 \\
6518.37 &       26.0    &       2.831   &       $-$2.58 \\
6533.93 &       26.0    &       4.558   &       $-$1.29 \\
6546.24 &       26.0    &       2.758   &       $-$1.71 \\
6551.68 &       26.0    &       0.990   &       $-$5.73 \\
6556.79 &       26.0    &       4.796   &       $-$1.61 \\
6569.21 &       26.0    &       4.733   &       $-$0.21 \\
6581.21 &       26.0    &       1.485   &       $-$4.73 \\
6591.31 &       26.0    &       4.593   &       $-$1.98 \\
6592.91 &       26.0    &       2.727   &       $-$1.51 \\
6593.87 &       26.0    &       2.433   &       $-$2.25 \\
6597.56 &       26.0    &       4.795   &       $-$0.98 \\
6609.11 &       26.0    &       2.559   &       $-$2.58 \\
6609.68 &       26.0    &       0.990   &       $-$5.78 \\
6633.41 &       26.0    &       4.835   &       $-$1.26 \\
6633.75 &       26.0    &       4.558   &       $-$0.71 \\
6634.11 &       26.0    &       4.795   &       $-$1.23 \\
6648.08 &       26.0    &       1.011   &       $-$5.79 \\
6665.43 &       26.0    &       1.557   &       $-$5.60 \\
6710.32 &       26.0    &       1.485   &       $-$4.81 \\
6713.74 &       26.0    &       4.795   &       $-$1.43 \\
6715.38 &       26.0    &       4.607   &       $-$1.54 \\
6716.24 &       26.0    &       4.580   &       $-$1.81 \\
6149.26 &       26.1    &       3.889   &       $-$2.66 \\
6247.56 &       26.1    &       3.892   &       $-$2.33 \\
6369.46 &       26.1    &       2.891   &       $-$4.09 \\
6385.45 &       26.1    &       5.553   &       $-$2.63 \\
6432.68 &       26.1    &       2.891   &       $-$3.63 \\
6456.38 &       26.1    &       3.903   &       $-$2.03 \\
6482.20 &       26.1    &       6.219   &       $-$1.67 \\
6516.08 &       26.1    &       2.891   &       $-$3.33 \\
\enddata

\end{deluxetable}
\clearpage

\tablenum{4}
\tablecolumns{14}
\tablewidth{0pt}

\begin{deluxetable}{cccccccccccccc}
\tabletypesize{\scriptsize}
\tablecaption{AGB stars in common with Lapenna et al. (2014)}
\tablehead{
\colhead{Star Name}     &
\colhead{T$_{\rm eff}$$_{\rm spec.}$} &
\colhead{log(g)$_{\rm spec.}$} &
\colhead{[Fe I/H]} &
\colhead{[Fe II/H]} &
\colhead{vt} &
\colhead{RV$_{\rm helio.}$} &
\colhead{Star Name}     &
\colhead{T$_{\rm eff}$$_{\rm phot.}$} &
\colhead{log(g)$_{\rm phot.}$} &
\colhead{[Fe I/H]} &
\colhead{[Fe II/H]} &
\colhead{vt} &
\colhead{RV$_{\rm helio.}$} \\
\colhead{Ours}  &
\colhead{(K)}      &
\colhead{(cgs)}      &
\colhead{}      &
\colhead{}      &
\colhead{km s$^{\rm -1}$}      &
\colhead{km s$^{\rm -1}$}      &
\colhead{L2014}      &
\colhead{(K)}      &
\colhead{(cgs)}      &
\colhead{}      &
\colhead{}      &
\colhead{km s$^{\rm -1}$}      &
\colhead{km s$^{\rm -1}$}
}

\startdata
2M00235852$-$7206177    &       4600    &       1.33    &       $-$0.78 &       $-$0.78 &       1.75    &       $-$1.03 &       100119  &       4500    &       1.65    &       $-$0.85 &       $-$0.71 &       1.80    &       $-$0.83 \\
2M00241142$-$7206126    &       4600    &       1.72    &       $-$0.72 &       $-$0.72 &       1.55    &       $-$20.75        &       100142  &       4550    &       1.70    &       $-$0.90 &       $-$0.80 &       1.60    &       $-$20.41        \\
2M00242763$-$7205213    &       4550    &       1.64    &       $-$0.67 &       $-$0.67 &       1.70    &       $-$20.96        &       200021  &       4550    &       1.70    &       $-$0.83 &       $-$0.84 &       1.90    &       $-$21.07        \\
2M00242846$-$7205019    &       4625    &       1.65    &       $-$0.76 &       $-$0.76 &       1.75    &       $-$22.21        &       200023  &       4575    &       1.75    &       $-$0.81 &       $-$0.79 &       1.80    &       $-$22.19        \\
\enddata

\end{deluxetable}
\clearpage
\end{landscape}

\end{document}